\documentclass[12pt]{iopart}
\usepackage{bm}% bold math
\usepackage{amssymb}
\usepackage{amsfonts}
\usepackage{graphicx}
\def\simleq{\; \raise0.3ex\hbox{$<$\kern-0.75em \raise-1.1ex\hbox{$\sim$}}\; }
\def\simgeq{\; \raise0.3ex\hbox{$>$\kern-0.75em \raise-1.1ex\hbox{$\sim$}}\; }
\def\arcsec{$''$}
\def\arcmin{$'$}
\newcommand{\eV}{{\rm eV}}

\newcommand{\GeV}{{\rm GeV}}
\newcommand{\TeV}{{\rm TeV}}
\newcommand{\PeV}{{\rm PeV}}

\newcommand{\pc}{{\rm pc}}

\newcommand{\muG}{\mu{\rm G}}
\newcommand{\mG}{{\rm mG}}
\newcommand{\s}{{\rm s}}

\newcommand{\Mpl}{M_{\rm Pl}}

\begin{document}
%opening
%  Two possibilities (JK)
%\title{New Crab-Nebula constraints on Planck-scale Lorentz Violation in QED}
\title{New constraints on Planck-scale Lorentz Violation in QED from
the Crab Nebula}
%: an {\it ab initio} computation and observational constraints}
\author{Luca Maccione$^{1,2}$, Stefano Liberati$^{1,2}$, Annalisa
Celotti$^{1,2}$, John G. Kirk$^3$} \address{$^1$ SISSA, Via Beirut, 2-4,
I-34014, Trieste, Italy} \address{$^2$ INFN, Sezione di Trieste, Via
Valerio, 2, I-34127, Trieste, Italy} \address{$^3$ Max-Planck-Institut
f\"ur Kernphysik, Saupfercheckweg, 1, D-69117, Heidelberg, Germany}
\eads{\mailto{maccione@sissa.it}, \mailto{liberati@sissa.it},
\mailto{celotti@sissa.it}, \mailto{John.Kirk@mpi-hd.mpg.de}}

\begin{abstract}
We set constraints on $O(E/M)$ Lorentz Violation in QED in an
effective field theory framework. A major consequence of such
assumptions is the modification of the dispersion relations for
electrons/positrons and photons, which in turn can affect the
electromagnetic output of astrophysical objects.  We compare the
information provided by multiwavelength observations with a full and
self-consistent computation of the broad-band spectrum of the Crab
Nebula.  We cast constraints of order $10^{-5}$ at 95\% confidence
level on the lepton Lorentz Violation parameters.
\end{abstract}
\section{Introduction}
\label{sec:Intro}

Local Lorentz Invariance (LI) is fundamental to both of the two pillars of
our present physical knowledge: the standard model of particle physics
and general relativity. Nonetheless, the most recent progress in
theoretical physics, in particular towards the construction of a theory
of Quantum Gravity (QG), has led to a new perspective, in which both of
the above mentioned theories are seen as effective ones to
be replaced by a theory of some more fundamental objects at high
energies. From this perspective it is conceivable that even
fundamental spacetime symmetries (such as local LI) could cease to be valid
in the vicinity of the Planckian regime.

This insight and several other arguments (see, e.g.,~the 
discussion in \cite{Mattingly:2005re, Jacobson:2005bg, AmelinoCamelia:2003uc, Smolin:2006pa}) that 
question LI at high energies, have been strengthened by
specific hints of Lorentz Violation (LV) arising from preliminary
calculations in various approaches to QG. Examples include string
theory tensor VEVs~\cite{KS89}, spacetime foam~\cite{GAC-Nat},
semiclassical spin-network calculations in Loop QG~\cite{LoopQG},
non-commutative geometry~\cite{Carroll:2001ws, Lukierski:1993wx, AmelinoCamelia:1999pm}, some brane-world
backgrounds~\cite{Burgess:2002tb} and condensed matter analogues of
``emergent gravity''~\cite{Analogues}.
Although none of these calculations proves 
that Lorentz symmetry breaking is a necessary feature of QG, they do
stimulate research aimed at understanding the possible measurable
consequences of LV.

In recent years, attempts to place constraints on high-energy
deviations from LI have mainly focused on modified dispersion relations for
elementary particles. In fact, in most of the above mentioned QG
models, LV enters through dispersion relations which can
be cast in the general form (it is assumed, for simplicity,
that rotational invariance is preserved and only boost invariance is
affected by Planck-scale corrections):
\begin{equation}%
E^2=p^2+m^2+f(E,p;\mu;M)\;,%
\label{eq:disprel}%
\end{equation}%
where we set the low energy speed of light $c=1$, $E$ and $p$ are the
particle energy and momentum, $\mu$ is a particle-physics mass-scale
(possibly associated with a symmetry breaking/emergence scale) and $M$
denotes the relevant QG scale.  Generally, it is assumed that $M$ is
of order the Planck mass: $M \sim M_{\rm Pl} \approx 1.22\times
10^{19}\;$GeV, corresponding to a quantum (or emergent) gravity
effect. The function $f(E,p;\mu;M)$ can be expanded in powers of the
momentum (energy and momentum are basically indistinguishable at high
energies, although they are both taken to be smaller than the Planck scale), and the lowest order LV terms ($p^2$ and $p^3$) have been
mainly considered \cite{Mattingly:2005re}.

At first sight, it appears hopeless to search for effects suppressed by the
Planck scale. Even the most energetic particles ever detected (Ultra
High Energy Cosmic Rays, see, e.g.,~\cite{Roth:2007in,
Abbasi:2007sv}) have $E \simleq 10^{11}~\GeV \sim 10^{-8} \Mpl$. 
However, as discussed below, even tiny corrections can be
magnified into a significant effect when dealing with 
high energies (but still well below the Planck scale), 
long distances of signal propagation, or peculiar reactions (see,
e.g.,~\cite{Mattingly:2005re}).

However, it must be stressed that most of the above cited effects
require a theoretical framework within which it is possible to justify the use
of modified dispersion relations of the form (\ref{eq:disprel}),
in order to calculate reaction rates and, more generally, to describe fully the particle
dynamics.

In this regard, two alternative approaches  have been proposed. On the one hand, one can interpret the
dispersion relation \eref{eq:disprel} as a by-product of an Effective Field Theory
characterised by Planck suppressed LV operators. On the other hand, such
a dispersion relation can be interpreted as the Casimir invariant of
some new relativity group (which would then incorporates two invariant scales, $c$ and $\Mpl$). The latter  approach goes under the name of ``doubly (or deformed)-special relativity'' (DSR) \cite{AmelinoCamelia:2000mn, AmelinoCamelia:2000ge}. While at the moment there are formulations in momentum space of this proposal, its formulation in coordinate space is still far from being complete as well as an effective quantum field theory implementation of this framework (if any, see e.g.~\cite{AmelinoCamelia:2002dx}). Lacking such an information about the dynamics implied by the theory it is nowadays very difficult to cast stringent constraints on the DSR framework, in particular the constraint that will be discussed in this work will not apply.

In this paper, we aim instead to improve the constraints obtained in the best
studied framework of Effective Field Theory with non-renormalisable
(i.e.,~mass dimension 5 and higher) LV operators
(see~\cite{Mattingly:2005re,Myers:2003fd} and references therein). To
this end, we compute, within our test theory, the broad band emission of 
the Crab Nebula (CN),
exploiting recent observational and theoretical improvements in our astrophysical knowledge of this object.

The outline of the paper is the following: in the next section we
present our theoretical framework.  In Section~3 the current status of
the constraints on this test theory is reviewed, briefly describing
the main processes involved and the role played so far by the CN. In
Section~4 we discuss the current observations of the CN and their
theoretical interpretation. Section~5 studies the foreseeable role of
the departures from LI in our theory and the processes which are
included in the reconstruction of the CN spectrum (from radio to
$\gamma$-rays). Finally, in Section~6, we discuss the constraints that
observations place on the theory and the possible improvements
expected from future experiments like GLAST~\cite{glast}.

\section{QED with Lorentz Violations at order $O(E/M)$}
\label{sec:LVtheories}

Myers \& Pospelov \cite{Myers:2003fd} found that there are essentially
only three operators of dimension five, quadratic in the fields, that
can be added to the QED Lagrangian preserving rotation and gauge
invariance, but breaking local LI\footnote{Actually these criteria
allow the addition of other (even CPT) terms, but these would not lead
to modified dispersion relations (they can be thought as extra
interaction terms) \cite{Bolokhov:2007yc}.}. These terms, which result
in a contribution of $O(E/M)$ to the dispersion relation of the
particles, are the following:
\begin{equation}
-\frac{\xi}{2M}u^mF_{ma}(u\cdot\partial)(u_n\tilde{F}^{na}) + \frac{1}{2M}u^m\bar{\psi}\gamma_m(\zeta_1+\zeta_2\gamma_5)(u\cdot\partial)^2\psi\:,
\label{eq:LVterms}
\end{equation}
where $\tilde{F}$ is the dual of $F$ and $\xi$, $\zeta_{1,2}$ are
dimensionless parameters. All these terms also violate the CPT
symmetry (note that while CPT violation implies LV the converse is not
true \cite{Greenberg:2002uu}).

From \eref{eq:LVterms} the dispersion relations of the fields are
modified as follows. For the photon 
\begin{equation}
\omega_{\pm}^2 = k^2 \pm \frac{\xi}{M}k^3\:,
\label{eq:disp_rel_phot}
\end{equation}
(the $+$ and $-$ signs denote right and left circular polarisation), while
for the fermion (with the $+$ and $-$ signs now denoting positive and
negative helicity states% 
)
\begin{equation}
E_\pm^2 = p^2 + m^2 + \eta_\pm \frac{p^3}{M}\;,
\label{eq:disp_rel_ferm}
\end{equation}
with $\eta_\pm=2(\zeta_1\pm \zeta_2)$.  For the antifermion, it can be
shown by simple ``hole interpretation" arguments that the same
dispersion relation holds, with $\eta^{af}_\pm = -\eta^f_\mp$ where
$af$ and $f$ superscripts denote respectively anti-fermion and
fermion coefficients~\cite{Jacobson:2005bg,Jacobson:2003bn}.

The modified dispersion relations affect standard
processes (such as threshold reactions) and permit new processes. 
Of course, this
new physics will be visible only at sufficiently high energies, given
that Planck suppression characterises the new relations. To estimate
the energy scale, consider, for example, a threshold
reaction involving photons and electrons such as  photon-photon
annihilation, leading to electron-positron pair creation. In such a
case the characteristic energy scale of the process -- to which the LV
term $\xi k^3/M$ should be comparable -- is the electron mass $m_{\rm
e}$. Assuming a LV coefficient $O(1)$ the effect of the LV should
become visible around the critical energy $k_{\rm cr}\approx 10$
TeV. While prohibitive for laboratory experiments, this energy is
within the range of the observed phenomena in high energy
astrophysics.

Observations involving energies of the order of 10 TeV can thus
potentially cast an $O(1)$ constraint on the coefficients defined
above.  But what is the theoretically expected value of the LV
coefficients in the dispersion relations (\ref{eq:disp_rel_phot},
\ref{eq:disp_rel_ferm})?  In particular, renormalisation group effects
could, in principle, strongly suppress the low-energy values of the LV
coefficients even if they are $O(1)$ at high energies.  Let us,
therefore, consider the evolution of the LV parameters.

Bolokhov \& Pospelov \cite{Bolokhov:2007yc} recently addressed the problem of
calculating the renormalisation group equations for QED and the
Standard Model extended with dimension-five operators that violate
Lorentz Symmetry. For the case of QED, an approach similar to that of
Myers \& Pospelov \cite{Myers:2003fd} was taken.  In this context, the
evolution equations for the LV terms in Eq.~(\ref{eq:LVterms}) that produce
modifications in the dispersion relations, can be inferred to be
\begin{equation}
\fl
\label{eq:RG}
\frac{d\zeta_1}{dt} =  \frac{25}{12}\,\frac{\alpha}{\pi}\,\zeta_1\; , \quad
\frac{d\zeta_2}{dt} =  \frac{25}{12}\,\frac{\alpha}{\pi}\,\zeta_2 - \frac{5}{12}\,\frac{\alpha}{\pi}\,\xi\; , \quad
\frac{d\xi}{dt} =  \frac{1}{12}\,\frac{\alpha}{\pi}\,\zeta_2 - \frac{2}{3}\,\frac{\alpha}{\pi}\,\xi \;,
\end{equation}
where $\alpha = e^2/4\pi \simeq 1/137$ ($\hbar = 1$) is the fine
structure constant 
and $t = \ln(\mu^2/\mu_0^2)$ with $\mu$ and $\mu_0$
two given energy scales. (Note that the above formulae are given to lowest order in powers of the electric charge, which allows one to neglect the 
running of the fine structure constant.)

These equations show that the running is only logarithmic and
therefore low energy constraints are robust. Moreover, they also show
that $\eta_{+}$ and $\eta_{-}$ cannot, in general, be equal at all
scales.

\section{Previous astrophysical constraints on QED with $\mathbf{O(E/M)}$ LV}
\label{sec:otherconstr}

\begin{figure}[htb]
\centering
\includegraphics[scale = 0.5]{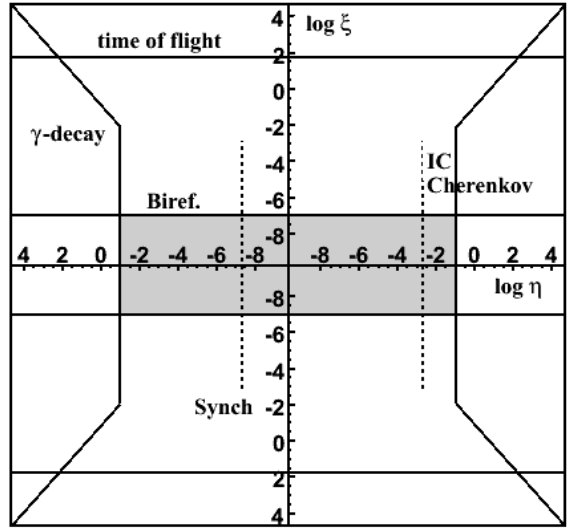}
\caption{Present constraints on the LV coefficients for QED with
dimension 5 Lorentz violation. The grey area is the allowed one and,
within it, the region bounded by the two dashed vertical lines
identifies the allowed range for at least one of the four lepton LV
coefficients (if one assumes that a single lepton population is 
responsible for the synchrotron and inverse Compton
emission of the CN simultaneously). }
\label{fig:constraints}
\end{figure}

Having addressed the issue of the natural values expected for
the LV coefficients, we summarise here the current status of the
astrophysical constraints on our test theory (see
\fref{fig:constraints}) and briefly explain how these constraints are
obtained.

\subsection{Photon time of flight}

Although these kinds of constraint are not directly relevant to our work, and
currently provide limits several orders of magnitude weaker than those
we shall present, we briefly discuss them, because they are widely
considered in the astrophysical community.

The dispersion relation \eref{eq:disp_rel_phot} implies that photons
of different colours (wave vectors $k_1$ and $k_2$) travel at slightly
different speeds. When accumulated over a cosmological distance $d$, this
effect produces a time delay
\begin{equation}
 \Delta t = \xi (k_2-k_1)d/M\;,
\end{equation}
which clearly increases with $d$ and with
the energy difference. The largest systematic error affecting this
method is the uncertainty about whether photons of different energy are
produced simultaneously in the source (e.g.,~a Gamma-Ray Burst, GRB).
One way of avoiding this problem, available only in the context of our test theory, would
be to measure the velocity difference of the two polarisation states
at a single energy, corresponding to
\begin{equation}
 \Delta t = 2|\xi|k\, d/M\;.
\end{equation}
However, this bound would require that both polarisations are observed
and that no spurious helicity dependent mechanism (such as, for example,
propagation through a birefringent medium) has affected the
relative propagation of the two polarisation states.

So far, the most robust constraints on $\xi$, derived from time of
flight differences, have been obtained from a statistical analysis
applied to the arrival times of sharp features in the intensity at
different energies from a large sample of GRBs with known
redshifts~\cite{GRBroboust}, leading to limits $\xi\leq
O(10^3)$. Using single objects (generally Blazars or GRBs) it is
possible to obtain a stronger, but less robust, constraint of order
$|\xi|\leq O(10^{2})$~\cite{TOF}.

\subsection{Birefringence}
According to Eq.~\eref{eq:disp_rel_phot} LV parameters for left and right circularly polarised photons are of opposite sign. As a consequence, during signal propagation, linear polarisation is rotated through an energy dependent angle. The
difference in rotation angle is
\begin{equation}
 \Delta\theta = \xi(k_2^2-k_1^2)d/2M\;.
\end{equation}
The fact that polarised photon beams are indeed observed from distant
objects imposes constraints on $\xi$. A strong constraint, $\xi
\simleq 2\times10^{-4}$, has been obtained by looking at UV radiation
from distant galaxies \cite{GK}. Recently, a claim of $|\xi| \simleq 2
\times 10^{-7}$ has been made using UV/optical polarisation measures
from GRBs \cite{Fan:2007zb}. This is currently the strongest
constraint on the LV coefficient for photons in the modified QED
considered here.

\subsection{Photon decay}
The decay of a photon into an electron/positron pair is made possible by 
LV because the
photon 4-vector is not null. This process has a threshold which,
if $\xi \simeq 0$, is set by the condition
\begin{equation}
k_{th} = (6\sqrt{3}m_e^2M/|\eta_\pm|)^{1/3}\;,
\end{equation}
where $\eta$ is the LV parameter for either an electron or a positron.
Since, from birefringence $\xi \simleq 10^{-7}$, the above expression
for the photon decay can be used to constrain the electron/positron
parameters. In \cite{Jacobson:2005bg} $|\eta_\pm| \simleq 0.2$ was
derived using the fact that 50~TeV $\gamma$-rays were measured from
the CN. This constraint has been tightened to $|\eta_\pm| \simleq
0.05$, thanks to observations of 80~TeV photons by HEGRA
\cite{Aharonian:2004gb}.

\subsection{Vacuum \v{C}erenkov -- IC electrons}
In the presence of LV the process of Vacuum \v{C}erenkov (VC)
radiation $e\rightarrow e\gamma$ can occur. Taking again $\xi \simeq
0$, the threshold energy is given by
\begin{equation}
 p_{\rm VC} = (m_e^2M/2\eta)^{1/3} \simeq 11~\TeV~\eta^{-1/3}\;.
\label{eq:VC_th}
\end{equation}
Moreover, just above threshold this process is extremely efficient,
having a time scale of order $\tau_{\rm VC} \sim 10^{-9}~\s$.  TeV
emission from the CN is usually attributed to the Inverse Compton
scattering (IC) of electrons/positrons on background photons (mainly
those from synchrotron radiation). These leptons would not be able to produce the
observed IC radiation if they were above the VC threshold, because
above $p_{\rm VC}$ the VC rate is much higher than the IC scattering
rate in the CN.

The observation of 50~TeV photons from the CN implies (by energy
conservation) the presence of at least 50 TeV leptons. This leads to
the bound $\eta \simleq 10^{-2}$ for at least one of the four fermion
parameters \cite{Jacobson:2005bg}. With the observation of 80~TeV
photons by HEGRA \cite{Aharonian:2004gb} the constraint can be
strengthened to $\eta \simleq 3\times10^{-3}$ as shown in
\fref{fig:constraints} (dashed vertical line in the positive $\eta$
range).

\subsection{Synchrotron radiation}
The synchrotron radiation emitted by electrons/positrons spiralling in a magnetic
field is strongly affected by LV. In the LI case, 
as well as in the LV one~\cite{Jacobson:2005bg,Jacobson:2002ye,Montemayor:2005ka}, most of the radiation from an electron of energy $E$ is emitted around a critical frequency
\begin{equation}
 \omega_c = \frac{3}{2}eB\frac{\gamma^3(E)}{E} 
% =\frac{3}{2}\frac{eB}{m_{\rm e}}\gamma^2\;,
\label{eq:omega_sync}
\end{equation}
where $\gamma(E) = (1-v^2(E))^{-1/2}$, and $v(E)$ is the electron
group velocity. 
However, in the LV case, the electron group velocity is given by
\begin{equation}
 v(E)= \frac{\partial E}{\partial p} =\frac{p}{E}\left(1+\frac{3}{2}\eta\frac{p}{M}\right)\,.
 % \simeq 1-\frac{m_e^2}{2E}+\eta\frac{E}{M} + O(\eta^2)\;.
\end{equation}
Therefore, $v(E)$ can exceed $1$ if $\eta > 0$ or can be strictly less
than the low energy speed of light if $\eta < 0$, resulting in $\gamma(E) \lessgtr E/m_e$ for
$\eta \lessgtr 0$. Moreover, if $\eta > 0$, $\gamma(E)$ grows
without bound until it diverges at  the soft VC threshold \eref{eq:VC_th},  which is well below the Planck scale for any $\eta\gg(m_e/M)^2\approx10^{-44}$.
On the other hand, for any $\eta < 0$, $\gamma(E)$ has a maximum and, hence, the critical frequency has an upper bound, $\omega_c^{\rm max}$.\footnote{
Notice that in our framework the sign of $\eta_{\pm}$ is undetermined. Conversely in DSR-like scenarios only superluminal parametrisation (i.e.~$\eta >0$) is allowed  \cite{AmelinoCamelia:2003ex} while in the string-inspired Liouville model of space-time foam \cite{Liouville} such coefficients are exactly zero and only the photon dispersion relation acquires a LV modification. This stresses the importance of a clear choice of framework when discussing this sort of phenomenological constraints (see e.g.~\cite{AmelinoCamelia:2002dx,Ellis:2003sd,Jacobson:2005bg}).}
Then, if synchrotron emission up to some maximal frequency $\omega_{\rm obs}$ is observed, one can deduce that the LV coefficient for the corresponding leptons cannot be more negative than the value for which $\omega_c^{\rm max}=\omega_{\rm obs}$. This leads to the bound~\cite{Jacobson:2005bg,Jacobson:2002ye}
\begin{equation}
\eta>-\frac{M}{m_e}\left(\frac{0.34\, eB}{m_e\,\omega_{\rm obs}}\right)^{3/2},
\end{equation}
which is strongest when the empirical ratio $B/\omega_{\rm obs}$ is minimised. Once again, the CN is the plerion for which the best
constraint can be cast.

Making the conservative assumption that the
100~MeV photons detected by $\gamma$-ray experiments \cite{EGRET} are
produced by synchrotron emission, a lower bound $\eta >
-8\times10^{-7}$, for at least one $\eta$, has been set
\cite{Jacobson:2005bg}. 
For the case of positive $\eta$, similar reasoning cannot be applied,
because for any positive $\eta$ a particle can emit all synchrotron
frequencies (up to infinity, in principle). Hence, a detailed reconstruction of the emitted spectrum is needed in this case.

\subsection{Helicity decay}
\label{subsec:HD}
Although it is not represented in \fref{fig:constraints}, this 
reaction is relevant to our investigation. In the
presence of LV, high energy electrons and positrons can flip their
helicity with the emission of a suitably polarised photon (Helicity
Decay, HD). This reaction does not have a real threshold, 
but rather an effective one \cite{Jacobson:2005bg}:
\begin{equation}
 p_{\rm HD} = (m_e^2M/\Delta\eta)^{1/3}\;,
\label{eq:HD_th}
\end{equation}
where $\Delta\eta = |\eta_+-\eta_-|$, at which the decay lifetime
$\tau_{HD}$ is minimised. For $\Delta\eta\approx O(1)$ this effective
threshold is around 10 TeV.  
For our purposes it is interesting to note that below threshold 
\begin{equation}
\tau_{\rm HD} > \Delta\eta^{-3} (p/10~\TeV)^{-8}\, 10^{-9}\s,
\label{eq:tauHD}
\end{equation}
while above threshold $\tau_{\rm HD}$ becomes independent of
$\Delta\eta$~ \cite{Jacobson:2005bg}. 
A constraint of $\Delta\eta < 0.4$ has been indirectly inferred in
\cite{Jacobson:2005bg} from the photon decay bound $|\eta_\pm| < 0.2$.

\subsection{An unfinished job}

We have seen that while the natural magnitude of the photon and
electron coefficients $\xi,\eta_\pm$ would be $O(1)$ if there were one
power of suppression by the inverse Planck mass, the coefficients are
currently restricted to the region $|\xi|\lesssim10^{-7}$ by
birefringence and $|\eta_{\pm}|\lesssim10^{-1}$ by photon decay.

Thus, whereas the constraint on the photon coefficient is remarkably
strong, the same cannot be said about the LV coefficients of
leptons. A constraint  on the lepton coefficients of comparable strength is
given by the synchrotron limit, but this is not double sided
and implies only that the LV coefficient of the population
responsible for the CN synchrotron emission cannot be smaller than
$-8\times10^{-7}$. Similarly the VC-IC bound $\eta<+3\times 10^{-3}$
constrains only one lepton population. These statements, although not
void of physical significance, cannot be considered constraints on
$\eta_\pm$, since for each of them one of the two parameters
$\pm\eta_+$ (and $\pm\eta_-$) will always satisfy the bound.

More can be said by using information obtained from current models
of the CN emission, in particular the fact that the CN emission is well 
fitted by assuming that
a single lepton population accounts for both the synchrotron and IC
radiation~\cite{Jacobson:2005bg}.  This implies that at least one of
the four pairs $(\pm\eta_\pm,\xi)$ must lie in the narrow region
bounded horizontally by the dashed lines of the synchrotron and IC
bounds and vertically by the birefringence constraint (see figure~1).
However, the dashed region limits apply at most to one of the four
pairs $(\pm\eta_\pm,\xi)$ as we cannot {\em a priori} exclude that
only one out of four populations is responsible for both the
synchrotron and IC emission (see again \cite{Jacobson:2005bg} for
further details).

It is clear that these simple arguments do not fully exploit the 
available astrophysical information. A detailed comparison of
the observations with the reconstructed spectrum in the LV case, where
all reactions and modifications of classical processes are considered,
can provide us with constraints on both positive and negative $\eta$
for the four lepton populations, at levels comparable to those already
obtained for the photon LV coefficient. Let us then move to reconsider
such information concerning the astrophysical object that so far has
proven most effective in casting constraints on the electron/positrons
LV coefficients: the CN.

\section{The Crab Nebula}
\label{sec:Crab}

The CN is a source of diffuse radio, optical and X-ray radiation
associated with a Supernova explosion observed in 1054~A.D., at a
distance from Earth of about $1.9\,$kpc.  A pulsar, presumably the
neutron star 
remnant of the explosion, is located at the centre of the Nebula, and
is believed to supply both the radiating particles (mostly electrons and positrons)
and magnetic fields, as well as the required power, which is somewhat 
less than the rotational energy loss-rate of the star (the 
\lq\lq spin-down luminosity\rq\rq), of roughly
$5\times10^{38}\,\textrm{erg/s}$ (for a recent review see
\cite{kirklyubarskypetri07}).

The Nebula emits an extremely broad-band spectrum (covering 21 decades in
frequency), produced by relativistic leptons via two major radiation
mechanisms: synchrotron radiation from radio to low energy
$\gamma$-rays ($E < 1\,\GeV$), and IC scattering 
for the higher energy $\gamma$-rays.  The
clear synchrotron nature of the radiation below $\sim 1$ GeV, combined
with a magnetic field strength of the order of $B \approx 100~\muG$
implies, when exact LI is assumed, the presence of relativistic
leptons with energies up to $10^{16}\,\eV$.  Their gyro period is
comparable to the synchrotron cooling timescale, implying an
acceleration rate close to the maximum estimated for shock-based
mechanisms (e.g.~\cite{achterbergetal01}).

The appearance of the CN depends on the observational wavelength: in
X-rays it is ellipsoidal, with angular dimensions $2'\,\times\,3'$
(corresponding to $\sim1.2\,\times1.8\,$pc at the Crab distance). In
the centre it displays a jet-torus structure in the X-ray and optical
bands. In the radio, this feature is less apparent, and the Nebula is
more extended, with a dimension of about
6\arcmin~\cite{Bietenholz:2004zg}.  Low frequency (radio) observations
allowed also the discovery of a number of structures on subarcsecond
scale, such as wisps, ripples, jets and arcs. Interestingly, most of
these substructures are seen also in the
optical \cite{hester} and  X-rays \cite{Mori:2004pt}.

In the frequency band ranging from radio to optical, the overall
emission spectrum of the CN has an extremely regular power-law shape,
with spectral index $\alpha_s = -0.27\pm0.04$ ($F\propto
E^{\alpha_s}$). The X-ray spectrum integrated over the whole Nebula 
is also well represented
by a power-law with spectral index $\alpha_s \approx -1.1$  in
the energy range 1-20 keV \cite{Seward:2006ff}.  In the 0.5-8 keV
band, the spectrum softens as the distance from the shock position in
the Nebula increases \cite{Seward:2006ff}, with the spectral index
passing from -0.9 in the inner region to about -2.0 in the outer
one. This result is confirmed by observations with the XMM-Newton
satellite \cite{Kirsch:2006yb}, and is consistent with the overall
expectations from the spherically symmetric magneto-hydro-dynamic
model by \cite{Kennel:1984vf} (although a detailed analysis reveals
effects that may be due to non-radial components of the flow velocity
\cite{morietal04}).

The $\gamma$-ray flux measured by EGRET \cite{EGRET} between 50~MeV
and 10~GeV matches well the extrapolation from the 50--500~MeV range
\cite{clear87}, which as been interpreted \cite{dejager92} as an
extension of the synchrotron spectrum from lower energies. The
observed hardening of the spectrum above 500~MeV was indeed predicted
by \cite{dejager92}, as due to the increased contribution of IC
emission.

Very high energy $\gamma$-rays from the CN were detected in the
pioneering observations by Whipple \cite{hillas98}, and, since then, by
several Imaging Atmospheric \v{C}erenkov Telescopes (IACT) (see, for example, \cite{Aharonian:2004gb, Kildea:2005ki, Wagner:2005pe,
Aharonian:2006pe}) and Extensive Air Shower (EAS) detectors.
The HEGRA stereoscopic IACTs observed the CN between 500 GeV and 80
TeV \cite{Aharonian:2004gb}.  The energy spectrum (with an overall
uncertainty of $\sim$15\%) is well approximated by a pure power-law with $\alpha_s\sim-1.62$.
%\begin{equation}
%\fl
%\frac{dF}{dE} = 
%(2.83\pm0.04\pm0.6)\times10^{-11}\,
%\left(\frac{E}{1~\TeV}\right)^{-2.62\pm0.02\pm0.05}\,
%\cm^{-2}\s^{-1}\TeV^{-1}\;.
%\end{equation}
Remarkably, the data show a 2.7$\sigma$
excess even at 86 TeV. Its position was determined to be shifted by
about 12\arcsec~(though with a systematic uncertainty of 25\arcsec) to
the west of the nominal position of the pulsar and consistent with the
centroid position of the X-ray emitting nebula.
The extension of the TeV excess (less than 3\arcmin~at 10 TeV)
excludes a strong contribution from the still undetected outer shock
of the expanding supernova remnant and is compatible with leptons
being accelerated in the proximity of the termination shock and then
cooled by the synchrotron/IC processes.  The data by HEGRA are also
compatible with the expected softening of the spectrum at high
energies ($E \geq 70~\TeV$).  This behaviour is confirmed by more
recent HESS observations \cite{Aharonian:2006pe}: the combined data
sets for the differential spectrum are best fitted by a power-law
(with slope $\alpha_s \sim -1.39$)
with an exponential cut-off at $14.3\pm2.1\pm2.8~\TeV$, rather than a
pure power-law.  Though the maximum lepton energy is model dependent,
the fact that photons with $E \simgeq 10~\TeV$ have been detected from
the CN is unambiguous evidence of the acceleration of
particles beyond $100~\TeV$.  Let us stress that this statement has to
be considered robust also in our test theory, in which
energy--momentum conservation still holds.

\subsection{Theoretical model}

From the theoretical point of view, the CN is one of the most studied
objects in astrophysics. However, in spite of more than 30 years of theoretical efforts,
important details of the interactions between the pulsar wind and the
synchrotron nebula are still missing.
The current understanding is based on a
spherically symmetric magneto-hydrodynamic (MHD) model presented in
two seminal papers by Kennel \& Coroniti \cite{Kennel:1984vf,
Kennel:1984vf2}, that accounts for the general features seen in the
spectrum.  In their model, the synchrotron nebula is powered by the
relativistic wind of cold electrons generated by the pulsar and
terminated by a standing reverse shock wave at $r_S \simeq 0.1~\pc$
\cite{Rees:1974nr}, due to the balance of the ram pressure of the flow
with the pressure of the outer nebula.
Kennel \& Coroniti found a stationary solution in which the particle flow evolves
adiabatically in the magnetised nebula.
The magnetisation of the flow is parametrised by $\sigma =
B^2/4\pi\gamma\rho$, which is the ratio between the magnetic
and the kinetic energy density in the flow ($\rho$ being the density).
The value of $\sigma$ is determined by the conditions at the outer
boundary of the Nebula. In fact, since the postshock flow has to match
the velocity of the material at the interface between the Nebula and
the remnant, which is non-relativistic, and a lower bound to the flow
speed is given by $v \simgeq \sigma / (1+\sigma)$
\cite{Kennel:1984vf}, it turns out that $\sigma$ has to be small, of
order $10^{-3}$ \cite{emmeringchevalier87}.  However, the magnetic
field is clearly dynamically important in the Nebula, because it is
responsible for its ellipsoidal shape
\cite{begelmanli92,vanderswaluw03}. Therefore, the value of $\sigma$
cannot be substantially less than $10^{-3}$.

\begin{figure}[htb]
 \centering
 \includegraphics[scale=0.5]{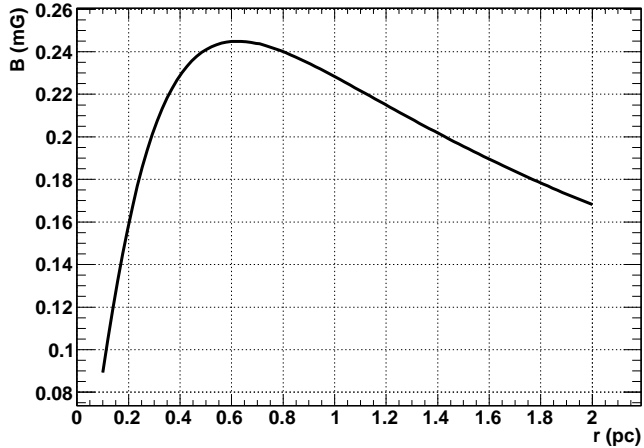}
 \caption{Magnetic field profile following Kennel \& Coroniti,
 assuming the best fit value $\sigma = 0.003$ and the reverse shock
 wave position $r_{\rm S}\approx 0.1$ pc.}
 \label{fig:magn_field}
\end{figure}

Observations of the radio and optical brightness distributions confirm
that the magnetic field, which is to a good approximation toroidal, is not
constant in the CN, but increases in the central region ($r \leq
0.5~\pc$) with distance from the pulsar. Its behaviour as a function of radius
downstream of the shock front 
(shown in \fref{fig:magn_field}) is determined initially by  
the gas pressure, which remains almost constant.
Because the gas behaves almost adiabatically, the density is 
also initially constant, and the radial velocity decreases 
as $1/r^2$, leading to an increase in the frozen in toroidal
magnetic field $B(r)$.
When $B(r)$ approaches the
equipartition value of $0.3~\mG$~\cite{marsden84} and starts to 
play a role in the dynamics of the flow, it ceases to grow and 
then falls off with increasing distance.

\section{The Crab model revisited}
\label{sec:revisited}

Because we consider a LV version of electrodynamics, we must check
whether this introduces modifications into the model of the CN and, if
so, what effects it produces.

The observed spectrum of the CN is the composite result of several
processes, as sketched in sec.~\ref{sec:Crab}. 
The way LV affects the physics of the CN is basically twofold. On the
one hand, classical processes, such as acceleration, synchrotron
emission and IC scattering, can be modified. On the other hand, new
processes (such as VC or HD) come into play.  On general grounds, we
expect both the modifications and the new processes to be important at
energies above $(m_e^2M/\eta)^{1/3} \approx 10~\TeV \eta^{-1/3}$,
which is the typical scale of threshold for almost all LV features. We
now investigate how the processes at work in the CN would appear in a
LV framework.

\subsection{Acceleration}
Several mechanisms have been suggested for the formation of the
spectrum of energetic leptons in the CN.  The spectrum is unusual in
the sense that most of the leptons are concentrated at low energy,
$E\sim 100\,$MeV, whereas the energy density contained in each decade
of particle energy peaks at about $1\,$TeV. Above this energy, the
spectrum appears to fall off roughly as $E^{-2.2}$. Most of the
uncertainty concerning the acceleration mechanism refers to the low
energy part of this distribution, $E\le1\,$TeV. The power-law spectrum
of high energy particles is usually interpreted in terms of the first
order Fermi mechanism operating at the ultra-relativistic termination
shock front of the pulsar wind, since, in the simplest kinematic
picture, this mechanism predicts a power law index of just the right
value \cite{kirklyubarskypetri07}.
 
In this picture, electrons and positrons acquire energy by crossing
and recrossing a shock front that propagates in a magnetised
medium. In between crossings, they are continually deflected by the
random magnetic field present in the medium, and may, therefore,
reverse their direction of travel. The magnetic deflections do not
change the particle energy as seen from a reference frame travelling
with the plasma. However, when a particle crosses a shock, it is exposed to
magnetic fluctuations embedded in an approaching plasma flow. These
increase the particle energy whenever it is deflected back
to the shock front.

The cycle of shock crossing and recrossing can be accomplished many
times, and each time the particle has a finite probability of escaping
from the vicinity of the shock front into the downstream medium, never
to return. The competition between energy gain and escape leads to a
scale-free power-law spectrum of accelerated particles. The power-law
index depends on the shock compression ratio, and, in the relativistic
case, on the angular dependence of the deflection process. In the
ultra-relativistic case, a variety of scattering laws have been tested
by different methods \cite{kirk05,morlinoetal07}, all of which appear
to give an index in the range $-2$ to $-2.3$, close to the asymptotic
value of $-2.23$, which can be derived semi-analytically for isotropic
diffusion in angle \cite{kirketal00}.

\subsubsection{The Fermi mechanism in the LV scenario.}
From the LV point of view, the important issue about the Fermi
mechanism is that its scale-free nature rests purely on the angular
distribution of the particles at the shock front. This determines both
the escape probability $P_{\rm esc}$ of a particle that crosses from
upstream to downstream, as well as the average change in particle
Lorentz factor $\left<\Delta\gamma\right>$. The latter is found by
convolving the Lorentz boost into the upstream plasma frame with the
return boost, averaged in each case over the angular distribution
function. Both $P_{\rm esc}$ and $\left<\Delta\gamma\right>$ are
independent of the length scale associated with the scattering process
-- an increase in the scattering mean free path simply produces a
longer time interval between crossings, and a slower fall-off of the
particle distribution with distance from the shock, but changes
neither the angular distribution at the shock front, nor the escape
probability.

In the LV picture, both the particle energy and Lorentz factor enter
into the computation of the trajectory. But, given that the angular
distribution is not a function of either of these, the spectrum
produced by the first order Fermi mechanism depends on the Lorentz
factor alone, through the quantity $\left<\Delta\gamma\right>$. On the
other hand, the maximum energy $E_{\rm c}$ to which the process can
accelerate particles may depend on loss processes as well as on the
time interval between shock crossings, which controls the acceleration
rate.  In the standard LI case, $E_{\rm c}$ is essentially determined
either by setting the particle gyro radius to equal the size of the
system, or the acceleration rate (which scales with the gyro
frequency) to equal the loss rate. In each case, a condition on the
particle energy rather than the particle Lorentz factor results. In
the CN, if we phenomenologically model the cut-off at $E_{\rm c}$ as
an exponential, we expect a particle spectrum in the high energy
region, $E>1\,$TeV, of the form
\begin{eqnarray}
n(E)&\propto&\gamma(E)^{-p}\textrm{e}^{-E/E_{\rm c}}
\end{eqnarray}
with $p\approx 2.4$ and $E_{\rm c}\approx 2.5\times10^{15}\,$eV.\footnote{Super-exponential cut-off spectral shapes do not lead to
significant differences in the output spectrum. For this reason we
considered a simple exponential cut-off, which also gives the best
fit to the data.}

Then, we can safely deal with the electron/positron distributions
inferred by \cite{Kennel:1984vf2,Aha&Atoy}, paying attention to
replace the energy with the Lorentz boost factor in the expressions
given by \cite{Aha&Atoy}. Of course, as we mentioned, the cut-off of
the spectrum results in a condition on the particle energy rather
than its boost.

\subsubsection{The role of Vacuum \v{C}erenkov emission.}
\label{subsubsec:VC}

However, in the LV theory there are additional mechanisms that can
influence $E_{\rm c}$ because the modified dispersion relations that we
consider allow processes that are otherwise forbidden. In particular, the VC
emission, due to its extreme rate above threshold, can produce a sharp
cut-off in the acceleration spectrum (let us remind that above
threshold $\tau_{VC}\sim 10^{-9}~\s$, to be compared with the
acceleration time scale $\tau_{acc}> 10^{3}~\s$).\footnote{An order of
magnitude estimate of a lower limit to the acceleration timescale can
be obtained by assuming the particle doubles its energy whenever it
completes a cycle of crossing and recrossing the shock front.  Since
magnetic fields bend the particle trajectory to make it do this, a
lower limit to the cycle time is given by the gyro period (magnetic
turbulence may make the particle diffuse, enhancing the time needed to
complete a cycle). The acceleration timescale is, therefore, $\tau_a >
\gamma m_e /eB \approx 1.1\times 10^3~\s\,
\left(E/1~\TeV\right)\left(B/100~\muG\right)^{-1}$.}

One might expect that the VC radiation emitted by particles above
threshold should produce some modification in the spectrum. However,
since the photon LV parameter $\xi$ has been independently constrained
to be very small, the VC process occurs in the soft regime
\cite{Jacobson:2005bg}. In this regime the emitted photon carries
away a small fraction of the electron energy, being at most in the
optical/UV range. Moreover, the emitting leptons are just in the high
energy tail of the spectrum, so that they are few in number, compared
to the optical/UV emitting ones. Therefore, the contribution of VC to
the CN spectrum in the optical/UV range should be negligible.

\subsubsection{The role of Helicity Decay and spin precession.}

A more subtle effect in the determination of the emitting particle
spectrum is given by the HD process.  At high energies, the electron
and positron states have well-defined helicity and the LV coefficients
$\eta_\pm$ are different depending on the particle helicity.

As discussed, the HD rate peaks at energies around $p_{\rm HD}$ (see
\eref{eq:HD_th}). Below $p_{\rm HD}$ the rate increases with energy
and depends on $\Delta\eta$, while above $p_{\rm HD}$ it decreases
independently of $\eta$~\cite{Jacobson:2005bg}. The expressions
\eref{eq:VC_th} and \eref{eq:HD_th} for $p_{\rm VC}$ and $p_{\rm HD}$,
respectively, show that at most $p_{\rm HD} \simgeq p_{\rm VC}$ for
$\Delta\eta \lesssim \eta$ (otherwise $\Delta\eta\ll \eta$ and $p_{\rm
HD} \gg p_{\rm VC}$ ). Since the VC emission acts as a hard cut-off on
the accelerated particles, in our scheme the HD process will occur
only in the regime $p \lesssim p_{\rm HD}$, where the (not yet
maximised) rate grows with energy.

However, in order to understand whether the HD is effective, one has
to compare its typical time scale $\tau_{\rm HD}$, as given in
eq.~\eref{eq:tauHD}, with the other relevant ones. In particular, a
competitive process is the precession of the spin of a particle moving
in a magnetic field. According to \cite{book:Landau} in the LI case
(it can be checked that this is still valid in the LV case, without
any modification, to within $10^{-14}$) the rate of change of the spin
orientation with respect to the instantaneous direction of motion of
the lepton in the laboratory frame is~\footnote{A comment is in order
here as, in principle, there could be interference between spin
precession and HD. However, assuming a static magnetic field, during
spin precession the electron energy is constant while HD implies the
emission of a photon, leading to non conservation of the electron
energy. Therefore they cannot interfere. Moreover, if a constant
timelike vector $u^\mu$ is used to ``parametrise'' Lorentz symmetry
violation, a term like $\bar{\psi}\gamma_\mu u^\mu \psi$ can appear in
the Lagrangian, which mimics the usual electro-magnetic interaction
term. However, as long as $u^\mu$ is constant, it cannot give rise to
magnetic-like interactions.}
\begin{equation}
\frac{{\rm d}\theta}{{\rm d}t} \equiv \omega_{\rm SR} =
\frac{e}{m_e}\frac{g-2}{2}B\;,
\label{eq:motion_inter_theta}
\end{equation}
where $(g-2) = \alpha/\pi$ represents the anomalous magnetic moment of
the lepton and a constant magnetic field $B$ is assumed. (Indeed it is
expected that the magnetic field in the CN is constant only over some
typical correlation length after which it reverses sign. For our
purposes, however, it is the rate of spin-rotation that is relevant, 
rather than
its direction.)

The spin rotation will effectively prevent helicity decay if the
spin precession rate is faster than the time needed for LV induced
effect. Therefore, the HD will play a role in spite of the spin
precession whenever the condition
%\begin{equation}
$ \omega_{\rm SR}\tau_{\rm HD} \ll 1$
%\label{eq:HDvsSp}
%\end{equation}
is met. 

According to \eref{eq:tauHD} and \eref{eq:motion_inter_theta} this
condition translates into
\begin{equation}
 p^{\rm (eff)}_{\rm HD} \simgeq 930~\GeV
 \left(\frac{B}{0.3~\mG}\right)^{1/8}|\Delta\eta|^{-3/8}\;.
\label{eq:peffhd}
\end{equation}
Electrons and positrons with $E > p^{(\rm eff)}_{\rm HD}$, can therefore be
found only in the helicity state that corresponds to the
lowest value of $\eta_\pm$. Correspondingly, the population of greater
$\eta$ will be sharply cut off above $p^{\rm (eff)}_{\rm HD}$ while the other will
increase.

Finally, the HD process implies the emission of a suitably polarised
photon. However, this has no
consequences for the observed spectrum, since the mean photon energy
(for $\Delta\eta \sim 10^{-6}$) is well within the radio band, where
the synchrotron spectrum, emitted by an overwhelmingly large 
number of low energy
electrons, is dominant. We refer the reader to \ref{app:HD} for a detailed discussion.

\subsection{More on synchrotron radiation}
The main modifications of the synchrotron emission process in presence
of LV have already been presented in Section~\ref{sec:LVtheories}. Let
us however consider them in more detail.

\begin{figure}
 \centering
 \includegraphics[scale=0.5]{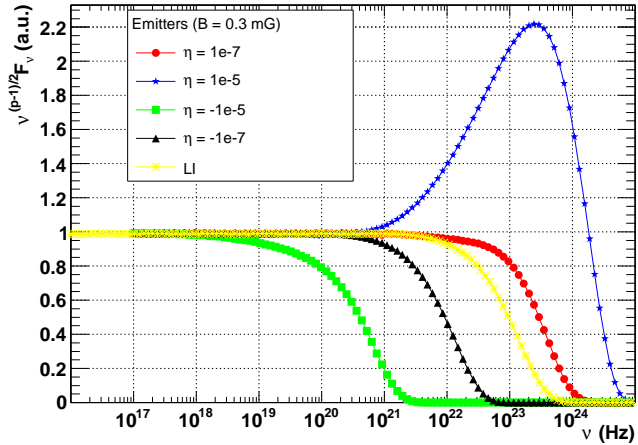}
 \caption{Modifications occurring due to LV in the synchrotron
 spectrum produced by a power-law distribution of leptons. The
 spectrum is normalised to a LI one without cut-off.}
 \label{fig:gr_show}
\end{figure}

There is a fundamental difference between particles with positive or
negative LV coefficient $\eta$. If $\eta$ is negative the group
velocity of the electrons is strictly lower than the (low energy)
speed of light. This implies that, at sufficiently high energy,
$\gamma(E)_{-} < E/m_e$ for all $E$.  As a consequence, the critical
frequency $\omega_c^{-}(\gamma, E)$ is always lower than the LI one,
and so the exponential cut-off of the LV synchrotron spectrum will be
at lower frequencies than in the LI case, as illustrated in
\fref{fig:gr_show}.

On the other hand, particles with a positive LV coefficient can be
superluminal and, therefore, $\gamma(E)$ increases more rapidly than $E/m_e$, reaching infinity at a finite energy, which corresponds to the threshold for
soft VC emission.\footnote{Of course such an infinity will be automatically ``regulated" by the fact that, as the electron approaches the threshold, its energy loss rate will at some point exceed its rate
of energy gain, thus preventing further acceleration.} Therefore, the critical frequency is also larger than
that found in the LI case, and the spectrum will show a characteristic
hump due to the enhanced $\omega_c$ (see ~\fref{fig:gr_show}).

The final remark concerns the characteristic synchrotron loss
timescale, defined as $\tau_{cool} = E/(dE/dt)$.  The classical result
for an electron spiralling in a given magnetic field is
\begin{equation}
 \frac{dE}{dt} =
 -\frac{2\,e^4}{3\,m_e^2}\gamma^2\,B^2v^2\sin^2\phi\,,
 \label{eq:synch-rate}
\end{equation}
where $\phi$ is the pitch angle
% of the electron direction of motion
%with respect to the magnetic field $B$, $v$ is its velocity and
and $\gamma = E/m_e$.  As the electron loses most of its energy at
frequencies around the critical one, a comparable expression for the
LV case can be written as
\begin{equation}
 \frac{dE}{dt} \sim \frac{\omega_c}{\Delta\,t} \sim \frac{\gamma^4}{E^2}\,,
\end{equation}
where we have used eq.~\eref{eq:omega_sync} and the fact that the typical emission time at ultra-relativistic energy is $\Delta t = 2\pi R(E)/\gamma$, with $R(E)=
E/eB$. The numerical factor in front of expression (21) can be fixed
by fitting the energy loss rate at low energies, where $\gamma =
E/m_e$.
  
%This energy loss rate lies below the upper limit on the
%acceleration rate discussed in \sref{subsubsec:VC}, and, therefore,
%does not necessarily prevent acceleration.

One might wonder if this modified rate could alter the
effectiveness of the acceleration mechanism in producing the highest
energy leptons. In fact, while for $\eta<0$ this is not an issue, for
$\eta>0$ one expects the rate to grow much faster than the LI one for
sufficiently high energies. Nonetheless it is easy to see that
appreciable deviations from the LI rate \eref{eq:synch-rate} occur at
energies $E \simgeq 8~\TeV /\eta^{1/3}$, i.e.~in the proximity of the
VC threshold. Therefore the effective cut-off on the spectrum of the
injected particles is not significantly lowered by the synchrotron
cooling in the LV case.

\subsection{IC radiation}
\label{subsec:IC}

The IC process is not strongly affected by LV. At the kinematic level all LV
terms intervene at a level of $< 10^{-11}$ at $E \lesssim 1~\PeV$,
whereas the cross section should be corrected by adding factors
proportional to $p^3/\Mpl$ and, therefore, the LV contribution is again
suppressed at the same level. Since the IC and synchrotron emission
mechanisms can be thought as being due to the scattering of a lepton
off a real or a virtual photon respectively \cite{Lieu:1993ve}, one
may wonder why the synchrotron is much affected by LV while the IC is
not.

The main difference between the two processes is that IC
scattering involves the interaction between a real lepton and a real
photon, whereas the synchrotron process involves a virtual photon of the
magnetic field in which the lepton is spiralling. In the former
case, the interaction is effective no matter what the photon energy
is. In the latter case, however, the reaction is more subtle.  An
electron spiralling in a static magnetic field can exchange momentum
but not energy with the field, since it is static. Moreover, the
exchanged momentum is such that the electron accelerates and describes a
spiral trajectory. Therefore, a synchrotron emitting electron does not
interact with all possible virtual photons, but only with those
that provide it with the required momentum transfer. In a sense, this is
a sort of a resonant process, where the resonance is dependent not
only on the electron energy, but also on its velocity.

Therefore, the admixture of dynamical and kinematic variables in the
synchrotron emission process makes it much more sensitive to LV compared 
IC scattering, where only energetic considerations matter: this is
another example of the fact that, in LV reasoning, velocity (or boost
in $\gamma$) and energy are not quite the same concept.

\section{Results}
\label{sec:constraints}

In order to constrain our test theory by exploiting the information 
contained in broad band observations of the CN, we adopt the following
strategy. 

First of all we construct a numerical algorithm\footnote{The code
is written in C++ and takes great advantage of many tools provided by
the ROOT package, see http://root.cern.ch} that calculates in full
generality, for any set of LV parameters, the synchrotron emission
from a distribution of leptons, according to the model of the CN
presented in \sref{sec:Crab} and taking into account all the processes
discussed in \sref{sec:revisited}. Of course, the LI model is
recovered by simply setting $(\eta_+, \eta_-) = (0,0)$. Then, we fix
most of the model parameters (magnetic field strength and particle
energy density) so as to match observations from radio to soft X-rays,
i.e.~in a regime where the LV terms here considered are not expected
to produce significant effects.

This procedure leads to model parameters (and a LI spectrum) which are
in agreement with those providing the best fit to the data in
\cite{Aha&Atoy}. We report here the most relevant parameters, namely
$p=2.4$ as the spectral index and $E_c = 2.5~\PeV$ as the high energy
cut-off of the freshly accelerated wind leptons.\footnote{Note
however, that in the LV case the cut-off energy of the injected
particles can be lowered by the VC process.}  The same parameters are
known to be able to reproduce the IC part of the CN spectrum in the LI
case \cite{Aha&Atoy}. Given that the IC reaction is basically
unaffected by LV, agreement with the high energy data will hold also
for non zero LV coefficients. Of course this also implies that, at
least with current data accuracy, the IC cannot be used to improve on
the constraints obtainable from the synchrotron part of the spectrum.

\subsection{Spectra}

\begin{figure}[!ht]
 \centering
 \includegraphics[scale=0.45]{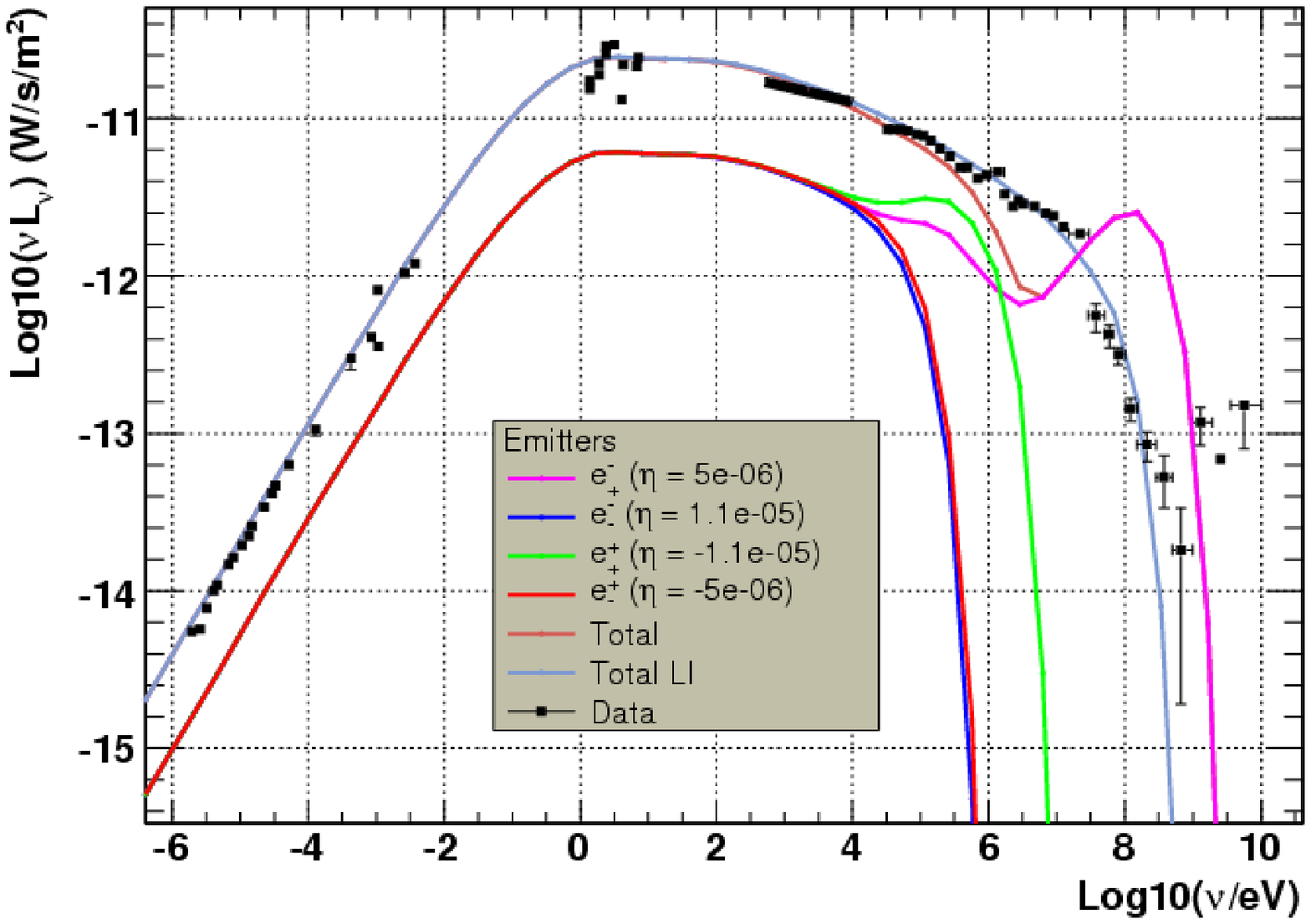}
 \includegraphics[scale=0.45]{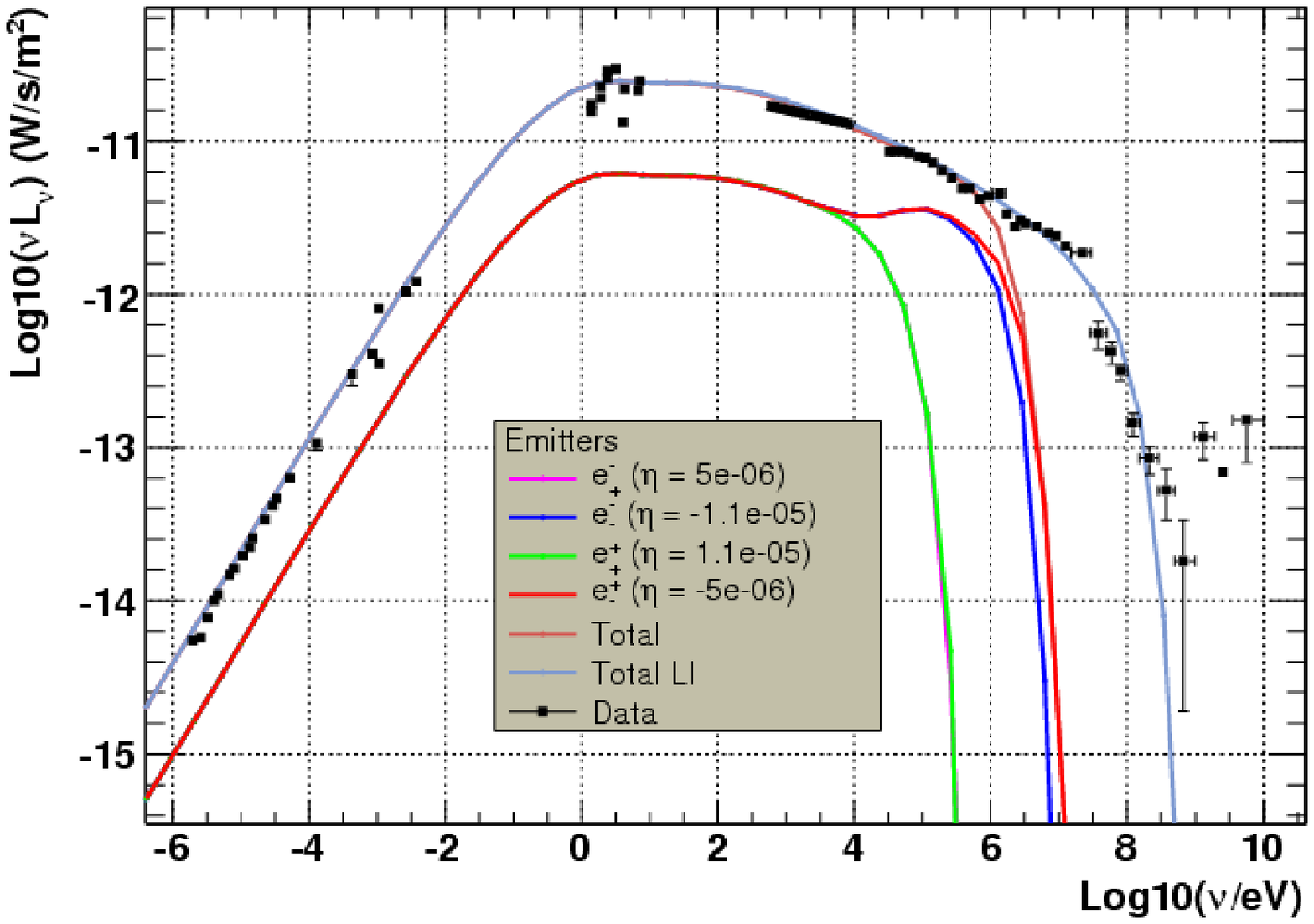}
 \caption{Comparison between observational data, the LI model and a LV
 one with $\eta_+\cdot\eta_- >0$ (left) and $\eta_+\cdot\eta_- <0$
 (right). The values of the LV coefficients are reported in the
 inserted panels and are chosen in order to show the salient features
 of the LV modified spectra. The leptons are injected according to the
 best fit values $p=2.4$, $E_c=2.5$ PeV. The individual contribution
 of each lepton population is shown.}
 \label{fig:gr_218}
\end{figure}

The general features of the spectra produced by our numerical
computation are illustrated in \fref{fig:gr_218} for
$\eta_+\cdot\eta_- >0$ (left panel) and $\eta_+\cdot\eta_- <0$ (right
panel) with $\eta_+$ assumed to be positive for definiteness. It is
clear that only these two cases are really different: in fact, the one
with both $\eta_\pm$ negative is the same as the $(\eta_+\cdot\eta_-
>0,\,\eta_+>0)$ case, while that with the signs scrambled is
equivalent to the case $(\eta_+\cdot\eta_- <0,\,\eta_+>0)$. This is
simply due to the fact that positron coefficients are related to
electron coefficients through $\eta^{af}_\pm = -\eta^{f}_\mp$ (see
section \ref{sec:LVtheories}).
%The values of $\eta_{\pm}$ are chosen in order to show the salient features of the LV modified spectra. 

%\begin{figure}[!ht]
% \centering \includegraphics[scale=0.4]{gr_90} 
% \caption{Comparison between observational data, the LI model and a LV
% one with $\eta_+\cdot\eta_- <0$.  The high energy emitters are produced according to the model of \cite{Aha&Atoy} with $p=2.4$, $E_c=2.5$ PeV. The individual contribution of each
% lepton population is shown in the inserted panel.}
% \label{fig:gr_90}
%\end{figure}

One can easily see that in the LI case the data are reasonably fitted
(as in \cite{Aha&Atoy}) and that the LV effects indeed appear at the expected
energy scales. Hence the procedure of fixing the free (LI) parameters
from the low energy observations is well defined.

The main difference between the left and right panels of
\fref{fig:gr_218} consists in the fact that in the first case
($\eta_+\cdot\eta_- >0$) only a population with positive $\eta$
survives to the HD, while in the opposite case ($\eta_+\cdot\eta_-
<0$) only a population with negative $\eta$ does. This has consequences
for the total synchrotron spectrum. In particular, the right panel of
\fref{fig:gr_218} shows a sharp cut-off since the high energy emission
in this case is produced by a population with negative $\eta$ which,
as discussed, has an upper bounded $\omega_c$. On the other hand, for
$\eta_+\cdot\eta_- >0$ (left panel of \fref{fig:gr_218}) a pronounced
feature appears with a dip followed by a hump. The dip is due to the
combination of two effects: the population is decaying with increasing
energy, while the critical frequency $\omega_c$ is growing faster than
``usual'' with energy. Hence, at some point the spectrum has a
minimum and then starts growing. Since, however, the population of the highest energy leptons
(responsible for the $\gamma$-ray part of the synchrotron spectrum) is
decaying very rapidly, the flux does, in the end, also decay. This
effect is responsible for the hump.

Finally, one might wonder if in this last case it is possible to
reproduce the high energy synchrotron emission even with $E_c \lesssim
1~\PeV$. However, this would require so high values of the LV
parameters (of order $10^{-4}\div10^{-3}$) that the resulting spectrum
would show a feature in hard-X/soft $\gamma$-rays incompatible with
the observations.  

\subsection{Constraints}

\begin{figure}[!ht]
 \centering
 \includegraphics[scale=0.5]{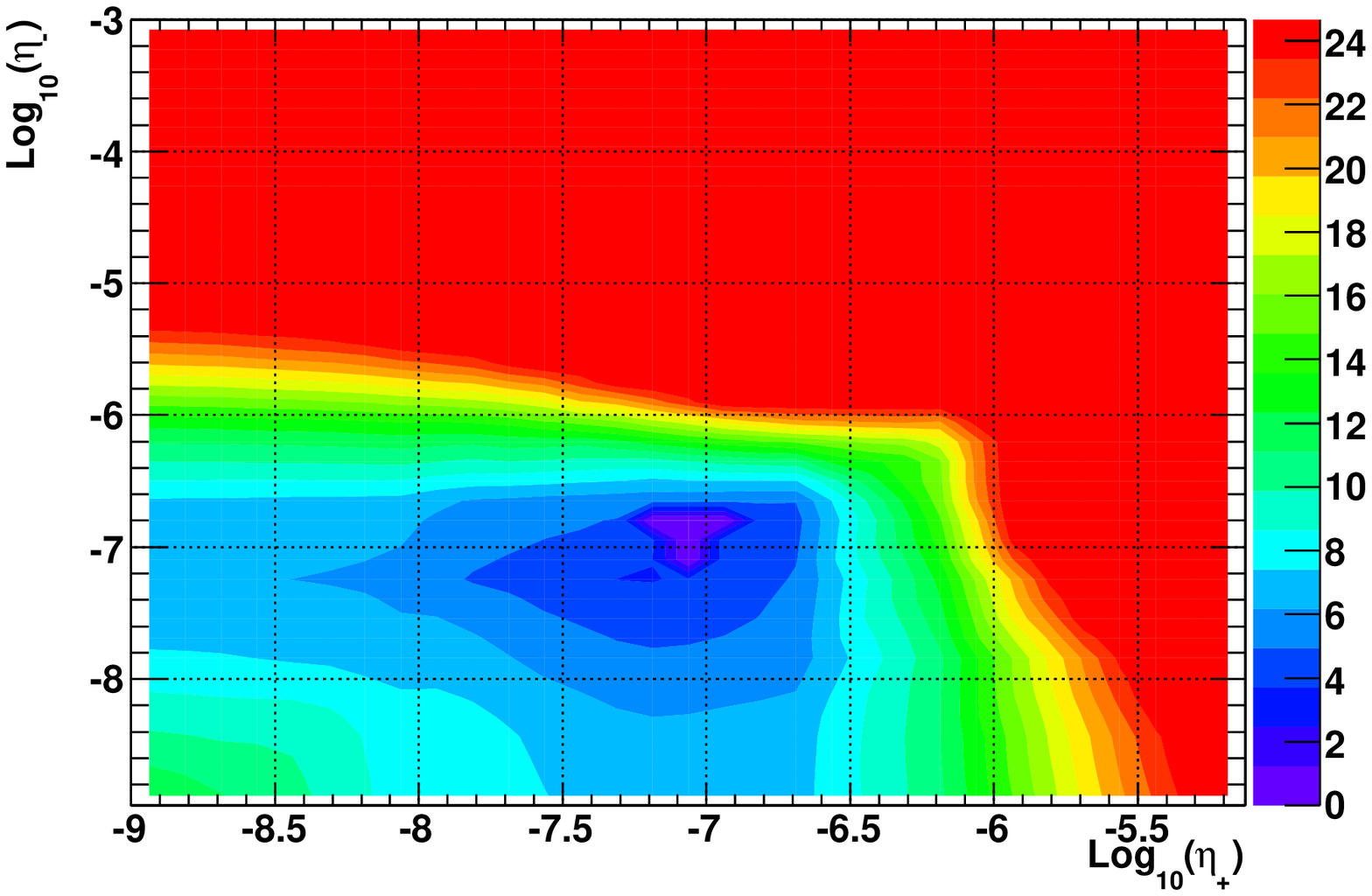}
 \caption{Contour plot of the reduced $\chi^2$ versus $\eta_+$ and
 $\eta_-$, in the case $\eta_+\cdot\eta_- >0$.}
 \label{fig:chi2_pos}
\end{figure}

\begin{figure}[!ht]
 \centering
 \includegraphics[scale=0.5]{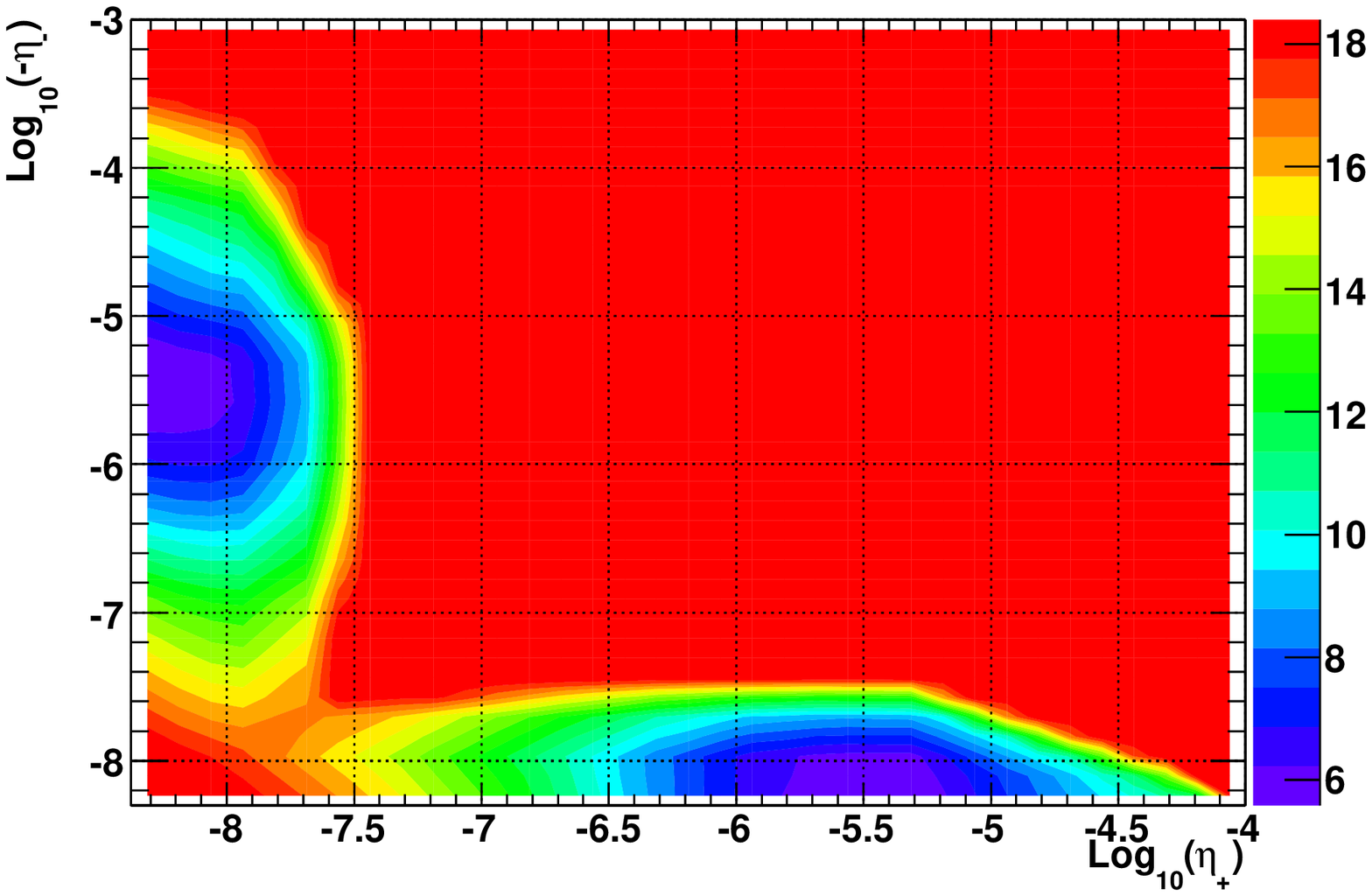}
 \caption{Contour plot of the reduced $\chi^2$ versus $\eta_+$ and
 $\eta_-$, in the case $\eta_+\cdot\eta_- <0$.}
 \label{fig:chi2_neg}
\end{figure}

In order to evaluate the constraints in an objective and quantitative manner, we present a $\chi^2$ analysis of the agreement between models and data.
\Fref{fig:chi2_pos} and \fref{fig:chi2_neg} show
the contour levels of
the reduced $\chi^2$ for the two 
cases $\eta_+\cdot\eta_- >0$ and $\eta_+\cdot\eta_-
<0$, respectively. Constraints at 90\%, 95\% and 99\%
Confidence Level (CL) correspond, respectively, to $\chi^2 > 8$, $\chi^2
> 10$, $\chi^2 > 13.5$. The minimum value of $\chi^2$ we obtain is
$\sim 3.6$ (see \cite{pdg} for more complete information).
From \fref{fig:chi2_pos} and \fref{fig:chi2_neg} we
conclude that the LV parameters for the leptons are both constrained,
at 95\% CL, to be $|\eta_\pm| < 10^{-5}$.

\begin{figure}[hbt]
 \centering
 \includegraphics[scale=0.5]{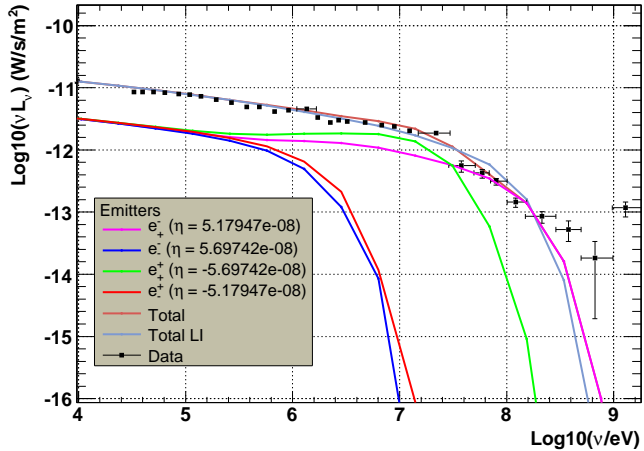}
 \caption{Best fit LV spectrum compared to the LI one.}
 \label{fig:goodfit}
\end{figure}

Our statistical analysis shows that there are values of the pair
$(\eta_+,\eta_-)$ that provide a better fit of the CN data than the LI
model. In particular, for $(\eta_+,\eta_-)\sim (5.2\times 10^{-8}, 5.7
\times 10^{-8})$ it is possible to reproduce (see \fref{fig:goodfit})
some features in the MeV range that are not found in the standard model.
Of course, while it is possible to explain these features by introducing new
components into the LI model, at the moment it seems that such alternatives would imply some sort of departure from the standard model of the CN emission (for example, \cite{Aha&Atoy} postulate
the existence of an additional population of emitting particles, with
a Maxwellian distribution).

Putting aside for the moment alternative (Lorentz invariant) models, something more can be said about the above result by  further investigating its statistical significance. This can be accomplished by assessing the significance of the difference between the  $\chi^2$ values of the best fit (LV) model and the standard LI one given that the extra two degrees of freedom characterising the LV case obviously allow for better fits. (Unfortunately, it is not possible to assess the probability to find the best fit value of the LV parameter $\eta$ given our ignorance of its theoretical expected magnitude or prior distribution.) 
This can be accomplished using the so called F-test \cite{ref:F-test}. We find a
value of 1.11 for the F-variable, from which we conclude that
the LI model and the best fit LV model are statistically
indistinguishable at 95\% CL. The critical value of the F-variable, for
which the models would indeed be distinguishable, is 1.67, and the
significant improvements in the $40-250$ MeV data expected
from the up-coming GLAST experiment may enable this value to be reached.

\section{Conclusions}

We have studied how relaxing the assumption of exact Lorentz invariance (within the
framework set up in \cite{Myers:2003fd}) influences the
electromagnetic output of astrophysical source models. In general, the most
important effects are those related to modifications of the particle
dispersion relations, which affect their propagation and their
interactions.

Starting from the most accurate theoretical model of the CN
\cite{Kennel:1984vf2,Aha&Atoy}, and taking into account the LV
contributions of {\em all} the electron/positron populations, we 
reproduce the observed synchrotron spectrum.  To do this, one must 
reconsider LI ``biases". Concerning the acceleration process,
we give arguments according to which the particle Lorentz boost,
rather than energy, enters in the acceleration spectrum. Moreover, we
study the effect of VC emission and HD on the emitting particle
distribution.

\begin{figure}
 \centering
 \includegraphics[scale=0.5]{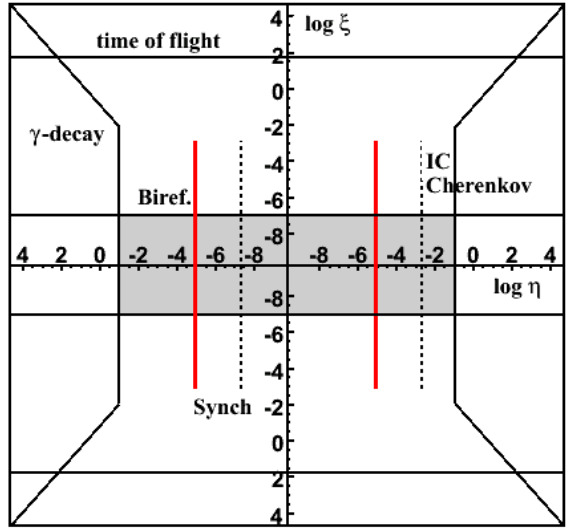}
 \caption{Updated overview of the constraints. The new allowed region of the parameter space is now the grey region bounded vertically by the birefringence constraint $\xi<O(10^{-7})$ and horizontally by the red lines representing the synchrotron constraint, $|\eta_\pm|<O(10^{-5})$, discussed here. 
 }
 \label{fig:new}
\end{figure}

The synchrotron, as well as the IC, processes are discussed and
the spectrum emitted by an arbitrary distribution of leptons, taking
into account all the subtleties occurring in LV reasoning, is
calculated numerically.  In this way {\em both} $\eta_\pm$ can be
constrained by comparing the simulated spectra to the observational
data. The $\chi^2$ statistics sets 90\% and 95\% CL exclusion limits
at $|\eta_\pm| < 10^{-6}$ and $|\eta_\pm| < 10^{-5}$,
respectively. The resulting state-of-the-art constraints are shown in
\fref{fig:new}.

The GLAST observatory is likely to achieve a significant 
step forward. An order of magnitude
estimate of the improvement can be obtained by considering its
sensitivity, which is $\sim$ 30 times better than that of EGRET in the relevant energy range. Assuming that GLAST will observe the CN at least as long as EGRET did (a very conservative assumption), measurement
errors in the 10 MeV-500 MeV band will be statistically reduced by
roughly a factor of $\sim$ 5. A constraint of order $10^{-6}$ at 99\%
CL ($10^{-7}$ at 95\% CL) would thus be within reach.

As a final remark, we would like to stress that the very tight
constraints achieved here on our test theory
\cite{Myers:2003fd,Jacobson:2003bn} show the remarkable potential of
this approach and suggest that similar studies should be undertaken
for other plausible theoretical frameworks.  For example, if one is
not willing to accept CPT violation in quantum gravity, then the QED
dimension 5 LV operators considered here would be forbidden and
dimension 6 operators (corresponding to $O(p^2/M^2)$ suppressed terms
in the dispersion relations) should be considered. In this direction,
on the one hand a theoretical development of the theory is much needed
as we still lack a formalisation of the QED extension in this case
(including the consideration of possible effects on lower dimensional
operators which could play a crucial role in casting
constraints). On the other hand, more accurate, higher energy and
possible new observations will be needed in order to overcome the
larger suppression of such higher order LV terms.

%It is thus
%auspicious that this kind of investigations will be pursued by a joint
%community of theorists and astrophysicists.

%unambiguously demonstrate that Planck-scale suppressed effects are definitely within the realm of current high energy astrophysics observations and, hence, that phenomenological studies of quantum gravity are already viable. Moreover, the fact that our test theory is now so constrained is 
\ack

We are grateful to Ted Jacobson and David Mattingly for many
fruitful discussions and the hints they gave us.  We also wish to
thank F. Aharonian and P. Blasi for some useful questions and remarks which helped
improving the clarity of the paper. The Italian MIUR is acknowledged
for financial support.

\appendix
\section{Comments on Helicity Decay}
\label{app:HD}

In order to understand what is the typical energy of the photon
emitted during the HD, we sketch here an estimate.  An incoming
electron with 4-momentum $p^\mu$ and LV parameter $\eta_1$
decays into another electron with 4-momentum $q^\mu$ and LV
coefficient $\eta_2$ plus a photon of 4-momentum $k^\mu$, whose angle
with respect to the direction of motion of the primary electron is
$\theta$. Since $\xi < 10^{-7}$ we set
$\xi = 0$.

Therefore, the dispersion relations for the particles involved are
\begin{equation}
 E_p^2 = m^2+p^2+\eta_1\frac{p^3}{M}\;,\quad
 E_q^2 = m^2+q^2+\eta_2\frac{q^3}{M}\;,\quad
 \omega^2 = k^2\;.
\label{eq:disp_hd}
\end{equation}

The conservation of energy-momentum $p^\mu = q^\mu+k^\mu$ implies,
when the ultra relativistic approximation is made for $E_q/q \approx
1+m^2/(2q^2)+\eta_2\,q/(2M)$, that
\begin{equation}
\fl
\eta_1\frac{p}{M} = \eta_2\frac{p}{M}\alpha(z,\theta)^{3}+2z\left(\alpha(z,\theta)+\frac{m^2}{2p^2\alpha(z,\theta)}+\eta_2\frac{p}{2\,M}\alpha(z,\theta)^2-\cos\theta+z\right)
\end{equation}
where $z=\omega/p$ and $\alpha(z,\theta) = \sqrt{1-2z\cos\theta+z^2}$.
If $z \ll 1$ (we will see that this assumption is justified {\em a
posteriori}),
\begin{equation}
 \Delta\eta \frac{p}{M} \approx z\left(\frac{m^2}{p^2}+2(1-\cos\theta)-3\eta_2\frac{p}{M}\cos\theta\right)\;,
\end{equation}
where $\Delta\eta \equiv \eta_1-\eta_2$ and whose solution is
\begin{equation}
 z(\theta, p) = \frac{\Delta\eta~ p^3/M}{m^2+2p^2(1-\cos\theta)-3\eta_2\,p^3/M\cos\theta}\,.
\label{eq:zeta_HD1}
\end{equation}

If far from the VC threshold \eref{eq:VC_th} (as in the cases
considered in this paper) one can safely neglect the last term in the
denominator of \eref{eq:zeta_HD1}, finding
\begin{equation}
 z(\theta,p) = \frac{\Delta\eta~ p^3/M}{m^2+2p^2(1-\cos\theta)}\,.
\label{eq:zeta_HD}
\end{equation}

From an experimental point of view, the angle $\theta$ is
unobservable. 
Hence, \eref{eq:zeta_HD} has to be averaged over the angular
distribution of the emitted photons. According to
\cite{Jacobson:2005bg}, we assume that, at lowest order, the matrix
element governing this process is angle independent.
Therefore, the mean photon energy is ($x \equiv \cos\theta$ and
redefine $z(\theta,p) \rightarrow z(\cos\theta, p)$)
\begin{equation}
\bar{z}(p) = \frac{\int_{-1}^1{\rm d}x~z(x,p)}{\int_{-1}^1{\rm d}x} =
\Delta\eta\frac{p}{2~M}\ln\left(2\frac{p}{m}\right)\;.
\label{eq:mean_ph_hd}
\end{equation}

Assuming a typical value $\Delta\eta = 10^{-6}$, from \eref{eq:peffhd}
$p_{\rm HD}^{\rm (eff)} = 160~\TeV$. \Eref{eq:mean_ph_hd} then implies
that $\bar{z}\left(p_{\rm HD}^{\rm (eff)}\right) = 1.7\times10^{-19}$,
{\em i.e.} $\bar{\omega} \simeq 2.8\times10^{-5}~\eV$, well within the
radio band.

%%%%%%%%%%%%%%%%%%%%%
\end{document}